\documentclass{aa}
\usepackage{graphicx}
\usepackage{bm}

\newcommand{\EQ}{\begin{equation}}
\newcommand{\EE}{\end{equation}}
\newcommand{\EQA}{\begin{eqnarray}}
\newcommand{\EEA}{\end{eqnarray}}
\newcommand{\brac}[1]{\langle #1 \rangle}
\newcommand{\pd}{\partial}

\newcommand{\DIV}{\vec{\nabla} \cdot }

\newcommand{\mean}[1]{\overline{#1}}
\newcommand{\meanv}[1]{\overline{\bm #1}}

\newcommand{\urms}{u_{\rm rms}}

\newcommand{\brms}{B_{\rm rms}}
\newcommand{\Beq}{B_{\rm eq}}

\newcommand{\kef}{k_{\rm f}}

\newcommand{\Sh}{{\rm Sh}}
\newcommand{\Pm}{{\rm Pm}}
\newcommand{\Rm}{{\rm Rm}}
\newcommand{\Pra}{{\rm Pr}}
\newcommand{\Ra}{{\rm Ra}}

\newcommand{\Co}{{\rm Co}}

\def\onethird{{\textstyle{1\over3}}}
\def\onehalf{{\textstyle{1\over2}}}
\def\threefourths{{\textstyle{3\over4}}}

\begin{document}

\authorrunning{K\"apyl\"a et al.}
\titlerunning{Large-scale dynamos in turbulent convection with shear}

   \title{Large-scale dynamos in turbulent convection with shear}

   \author{P. J. K\"apyl\"a
	  \inst{1}
          \and
          M. J. Korpi
          \inst{1}
          \and
          A. Brandenburg
	  \inst{2}
	  }

   \offprints{P. J. K\"apyl\"a\\
          \email{petri.kapyla@helsinki.fi}
	  }

   \institute{Observatory, T\"ahtitorninm\"aki (PO Box 14), FI-00014 
              University of Helsinki, Finland
         \and NORDITA, Roslagstullsbacken
              23, SE-10691 Stockholm, Sweden}

   \date{Received 2 June 2008 / Accepted 30 September 2008}

   \abstract{}%
   {To study the existence of large-scale convective dynamos
     under the influence of shear and rotation.}%
   {Three-dimensional numerical simulations of penetrative compressible convection
     with uniform horizontal shear are used to study dynamo action and
     the generation of large-scale magnetic fields.
     We consider cases where the magnetic Reynolds number is either
     marginal or moderately supercritical with respect to small-scale
     dynamo action in the absence of shear and rotation.
     Our magnetic Reynolds number is based on the wavenumber
     of the depth of the convectively unstable layer.
     The effects of magnetic helicity fluxes are studied by comparing
     results for the magnetic field with open and closed boundaries.
     }%
   {Without shear no large-scale dynamos are found even if the
     ingredients necessary for the $\alpha$-effect
     (rotation and stratification) are present in the system. When
     uniform horizontal shear is added, a large-scale magnetic
     field develops, provided the boundaries are open.
     In this case the mean magnetic field contains a significant fraction
     of the total field.
     For those runs where the magnetic Reynolds number is between 60 and 250,
     an additional small-scale dynamo is expected to be excited, but the field
     distribution is found to be similar to cases with smaller magnetic
     Reynolds number where the small-scale dynamo is not excited.
     In the case of closed (perfectly conducting) boundaries, magnetic
     helicity fluxes are suppressed and no large-scale fields are found.
     Similarly, poor large-scale field development is seen when vertical 
     shear is used in combination with periodic boundary conditions in the 
     horizontal directions.
     If, however, open (normal-field) boundary conditions are used 
     in the $x$-direction, a large-scale field develops.
     These results 
     support the notion that shear not only helps to generate the field, but
     it also plays a crucial role in driving magnetic helicity fluxes
     out of the system along the isocontours of shear, thereby allowing
     efficient dynamo action.
     }
 {}


   \maketitle


\section{Introduction}
It is evident that the Sun possesses a large-scale magnetic field. The
most likely origin of this field is a hydromagnetic dynamo working
within or somewhat below the turbulent convection zone
(e.g.\ Ossendrijver 2003). The solar dynamo is
generally thought to rely on large-scale shear to produce toroidal
magnetic field from poloidal field and a process called the
$\alpha$-effect which produces poloidal field from toroidal field.
In the turbulent
mean-field dynamo picture an $\alpha$-effect can arise due to
helical motions in a stratified fluid in the presence of either rotation
or shear, or both (Rogachevskii \& Kleeorin \cite{RogaKle2003};
R\"udiger \& Kitchatinov \cite{RueKit2006};
R\"adler \& Stepanov \cite{RaedStep2006}).
Other candidates for regenerating large-scale magnetic fields in shearing
turbulence include the incoherent $\alpha\Omega$-effect
(Vishniac \& Brandenburg \cite{VishBran1997}; Proctor \cite{Proctor2007})
and the shear--current effect (Rogachevskii \& Kleeorin \cite{RogaKle2003,RogaKle2004}).

Numerical simulations of convection have been used to
study dynamo action in local (e.g.\ Cattaneo \& Hughes
\cite{CattHugh2006}; Tobias et al.\ \cite{Tobiasea2008})
and global (e.g.\ Brun et al.\ \cite{BMT04}, Browning et al.\
\cite{Browning_etal06}, Brown et al.\ \cite{Brown_etal07}) settings.
So far, however, with the exception of the last two papers, these
models have had a hard time generating appreciable large-scale
magnetic fields, although all the ingredients such as
rotation and stratification necessary for an $\alpha$-effect to occur
have been present. Possible reasons for the lack of
large-scale dynamo action in the numerical models is arguably the
lack of scale separation between the energy carrying scale and the
scale of the domain and the absence of large-scale shear in the system.
In stratified convection, the lack of sufficient scale separation is difficult to tackle due to the
prohibitive computational challenges involved when trying to cover many more
vertical scale heights in the domain.
Adding shear, however, is easier to accomplish, which is one of the
principal aims of the present paper.

An additional point concerning large-scale dynamo action in turbulent
convection is the possibility of catastrophic quenching due to small-scale
magnetic fields. It has long been argued that small-scale fields cause
catastrophic quenching of the large-scale dynamo (e.g.\
Vainshtein \& Cattaneo \cite{VainCatt1992}). This sort of quenching, however, only
occurs in specific circumstances, i.e.\ when there is no magnetic
helicity flux out of the domain.
This behaviour can be understood in terms of magnetic helicity
(Brandenburg \& Subramanian \cite{BS05}) which is a
conserved quantity in ideal MHD. If magnetic diffusivity is
finite, the magnetic helicity can only change diffusively under the aforementioned
special circumstances that lead to resistively slow saturation of the large-scale
field.

If, however, magnetic helicity is driven out of the domain,
catastrophic quenching can be alleviated and the large-scale dynamo
is expected to saturate near equipartition field strengths.
A promising mechanism capable of driving magnetic helicity fluxes was 
introduced by Vishniac \& Cho (\cite{VishCho2001}), who found that in 
the presence of shear the flux follows isocontours of constant velocity
(Brandenburg \& Subramanian \cite{BS05b}; Subramanian \& Brandenburg \cite{SB06}).
We expect that this shear-driven magnetic 
helicity flux is of crucial importance for large-scale dynamo action.

The remainder of the paper is organised as follows: the numerical model 
is described in Sect.~\ref{sec:model}, and the results and conclusions 
are presented in Sects.~\ref{sec:results} and \ref{sec:conclusions}, 
respectively.

\section{The model}
\label{sec:model}

Our model setup is similar to that used by Brandenburg et al.\
(\cite{Brandea1996}), Ossendrijver et al.\ (\cite{Osseea2001})
and K\"apyl\"a et al.\ (\cite{Kaepylaeea2004},
\cite{Kaepylaeea2006}).
A rectangular portion of a star is modelled
by a box situated at colatitude $\theta$. The dimensions of the
computational domain are $(L_x, L_y, L_z) =
(4,4,2)d$, where $d$ is the depth of the convectively unstable layer,
which is also used as the unit of length. The box is divided into
three layers, an upper cooling layer, a convectively unstable layer,
and a stable overshoot layer (see below). The following set of
equations for compressible magnetohydrodynamics is being solved:
\begin{equation}
\frac{\mathcal{D} \bm A}{\mathcal{D}t} = -S A_y \hat{\bm{x}} - (\bm{\nabla}\bm{U})^{\rm T}\bm{A}-\mu_0\eta {\bm J}, \label{equ:AA}
 \end{equation}
\begin{equation}
\frac{\mathcal{D} \ln \rho}{\mathcal{D}t} = -\DIV{\bm U},
 \end{equation}
\begin{equation}
 \frac{\mathcal{D} \bm U}{\mathcal{D}t} = -SU_x\bm{\hat{y}}-\frac{1}{\rho}{\bm \nabla}p + {\bm g} - 2\bm{\Omega} \times \bm{U} + \frac{1}{\rho} \bm{J} \times {\bm B} + \frac{1}{\rho} \bm{\nabla} \cdot 2 \nu \rho \mbox{\boldmath ${\sf S}$}, \label{equ:UU}
 \end{equation}
\begin{equation}
 \frac{\mathcal{D} e}{\mathcal{D}t} = - \frac{p}{\rho}\DIV {\bm U} + \frac{1}{\rho} \bm{\nabla} \cdot K \bm{\nabla}T + 2 \nu \mbox{\boldmath ${\sf S}$}^2 + \frac{\mu_0\eta}{\rho} \bm{J}^2 - \frac{e\!-\!e_0}{\tau(z)}, \label{equ:ene}
 \end{equation}
where $\mathcal{D}/\mathcal{D}t = \pd/\pd t + (\bm{U} + \meanv{U}_0)
\cdot \bm{\nabla}$, and $\meanv{U}_0 = (0,Sx,0)$ is the imposed
large-scale shear flow. $\bm{A}$ is the magnetic vector potential,
$\bm{B} = \bm{\nabla} \times \bm{A}$ the magnetic field, and
$\bm{J} =\bm{\nabla} \times \bm{B}/\mu_0$ is the current density,
$\mu_0$ is the magnetic permeability, $\eta$ and $\nu$ are the magnetic diffusivity and
kinematic viscosity, respectively, $K$ is the heat conductivity, $\rho$ the density, $\bm{U}$ the
velocity, $\bm{g} = -g\hat{\bm{z}}$ the gravitational acceleration,
and $\bm{\Omega}=\Omega_0(-\sin \theta,0,\cos \theta)$ the rotation vector.
The fluid obeys an ideal gas law $p=\rho e (\gamma-1)$, where $p$
and $e$ are the pressure and internal energy, respectively, and
$\gamma = c_{\rm P}/c_{\rm V} = 5/3$ is the ratio of specific heats at
constant pressure and volume, respectively.
The specific internal energy per unit mass is related to the
temperature via $e=c_{\rm V} T$.
The rate of strain tensor $\mbox{\boldmath ${\sf S}$}$ is given by
\begin{equation}
{\sf S}_{ij} = \onehalf (U_{i,j}+U_{j,i}) - \onethird \delta_{ij} \DIV \bm{U}.
\end{equation}
The last term of Eq.~(\ref{equ:ene}) describes cooling at the top of
the domain. Here, $\tau(z)$ is a cooling time which has a profile
smoothly connecting the upper cooling layer and the convectively
unstable layer below, where $\tau\to\infty$.

The positions of the bottom of the box, bottom and top of the
convectively unstable layer, and the top of the box, respectively,
are given by $(z_1, z_2, z_3, z_4) = (-0.85, 0, 1, 1.15)d$. Initially
the stratification is piecewise polytropic with polytropic indices
$(m_1, m_2, m_3) = (3, 1, 1)$, which leads to a convectively unstable
layer above a stable layer at the bottom of the domain and an
isothermal cooling layer at the top.
All simulations with rotation use $\theta=0\degr$ corresponding to
the north pole.

\subsection{Nondimensional units and parameters}

Non-dimensional quantities are obtained by setting
\begin{eqnarray}
d = g = \rho_0 = c_{\rm P} = \mu_0 = 1\;,
\end{eqnarray}
where $\rho_0$ is the initial density at $z_2$. The units of length, time,
velocity, density, entropy, and magnetic field are
\begin{eqnarray}
&& [x] = d\;,\;\; [t] = \sqrt{d/g}\;,\;\; [U]=\sqrt{dg}\;,\;\; [\rho]=\rho_0\;,\;\; \nonumber \\ && [s]=c_{\rm P}\;,\;\; [B]=\sqrt{dg\rho
_0\mu_0}\;. 
\end{eqnarray}
We define the fluid and magnetic Prandtl numbers and the Rayleigh
number as
\begin{eqnarray}
\Pra=\frac{\nu}{\chi_0}\;,\;\; \Pm=\frac{\nu}{\eta}\;,\;\; \Ra=\frac{gd^4}{\nu \chi_0} \bigg(-\frac{1}{c_{\rm P}}\frac{{\rm d}s}{{\rm d}z
} \bigg)_0\;,
\end{eqnarray}
where $\chi_0 = K/(\rho_{\rm m} c_{\rm P})$ is the thermal
diffusivity, and $\rho_{\rm m}$ is the density in the middle of
the unstable layer. The entropy gradient, measured in the middle of
the convectively unstable layer in the non-convecting hydrostatic state,
is given by
\begin{eqnarray}
\bigg(-\frac{1}{c_{\rm P}}\frac{{\rm d}s}{{\rm d}z}\bigg)_0 = \frac{\nabla-\nabla_{\rm ad}}{H_{\rm P}}\;,
\end{eqnarray}
where $\nabla-\nabla_{\rm ad}$
is the superadiabatic temperature gradient with 
$\nabla_{\rm  ad} = 1-1/\gamma$, $\nabla = (\pd \ln T/\pd \ln
  p)_{z_{\rm m}}$, where $z_{\rm m}=z_3-z_2$, and $H_{\rm P}$ being the pressure scale height
(Brandenburg et al.\ \cite{Brandea2005}).
The amount of stratification is determined by the parameter 
$\xi_0 =(\gamma-1) e_0/(gd)$, which is the pressure scale height at
the top of the domain normalized by the depth of the unstable layer.
We use in all cases $\xi_0 =0.3$,
which results in a density contrast of about 23.
We define the magnetic Reynolds number via
\begin{eqnarray}
{\rm Rm} = \frac{\urms}{\eta \kef}\;,
\end{eqnarray}
where $\kef = 2\pi/d$ is assumed as a reasonable estimate
for the wavenumber of the energy-carrying eddies.
Note that our definition of $\Rm$ is smaller than the
usually adopted one by a factor $2\pi$.
The amount of shear and rotation is quantified by
\begin{eqnarray}
{\rm Sh} = \frac{S}{\urms \kef}\;,\quad
{\rm Co} = \frac{2\,\Omega_0}{\urms \kef}\;. \label{equ:ShCo}
\end{eqnarray}
The denominators in Eq.~(\ref{equ:ShCo}) give an estimate of the 
convective turnover time.
The equipartition magnetic field is defined by 
\begin{equation}
\Beq \equiv \langle\mu_0\rho\bm{U}^2\rangle^{1/2},\label{equ:Beq}
\end{equation}
where the angular brackets denote volume averaging.

\subsection{Boundary conditions}

\begin{table*}[t]
\centering
\caption[]{Summary of the runs. The numbers are given for the saturated state of the dynamo
(except for Run~E1, which was not run to saturation). Runs~D and D3 are the same model.
Here, $\mbox{Ma}=\urms/(gd)^{1/2}$,
$\tilde{B}_{\rm rms} \equiv \brms/B_{\rm eq}$,
$\brac{\tilde{\mean{B}}_y^2}^{1/2}=\brac{\mean{B}_y^2}^{1/2}/\brms$,
and $Q^{-1} = \brac{\mean{B}_x^2}^{1/2}/\brac{\mean{B}_y^2}^{1/2}$.
The last column gives the magnetic field boundary condition at the $z$-boundaries, except for the last two lines where the $x$-boundary condition is given. $\Pm=5$ in all models except D6 where $\Pm=10$.
}
      \label{tab:onlyshear}
      \vspace{-0.5cm}
     $$
         \begin{array}{p{0.035\linewidth}cccccccccccc}
           \hline
           \noalign{\smallskip}
Run & $grid$ & $\Pra$ & $\Ra$ & $Rm$ & \Sh & \Co & \mbox{Ma} & \tilde{B}_{\rm rms} & \brac{\tilde{\mean{B}}_y^2}^{1/2} & Q^{-1} & $BC$ \\ \hline 
A   & 256^3  & 0.69 & 6.1\cdot10^5 & 70  &     0  & 0    & 0.044 & 0.19 & 0.05 & 0.90 & $vf$ \\ 
B   & 256^3  & 0.69 & 6.1\cdot10^5 & 61  &     0  & 0.42 & 0.038 & 0.31 & 0.05 & 0.83 & $vf$ \\ 
C   & 256^3  & 0.69 & 6.1\cdot10^5 & 74  & -0.17  & 0    & 0.046 & 1.09 & 0.55 & 0.11 & $vf$ \\ 
C'  & 256^3  & 0.69 & 6.1\cdot10^5 & 92  & -0.14  & 0    & 0.058 & 0.37 & 0.22 & 0.31 & $pc$ \\ 
D   & 256^3  & 0.69 & 6.1\cdot10^5 & 56  & -0.23  & 0.46 & 0.036 & 1.41 & 0.61 & 0.15 & $vf$ \\ 
\hline
D1  &  64^3  & 2.74 & 1.5\cdot10^5 & 11  & -0.28  & 0.56 & 0.028 & 1.72 & 0.82 & 0.13 & $vf$ \\ 
D2  & 128^3  & 1.37 & 3.1\cdot10^5 & 25  & -0.25  & 0.50 & 0.032 & 1.45 & 0.67 & 0.17 & $vf$ \\ 
D3  & 256^3  & 0.69 & 6.1\cdot10^5 & 56  & -0.23  & 0.46 & 0.035 & 1.41 & 0.61 & 0.15 & $vf$ \\ 
D4  & 384^3  & 0.48 & 8.8\cdot10^5 & 83  & -0.22  & 0.44 & 0.036 & 1.37 & 0.58 & 0.16 & $vf$ \\ 
D5  & 512^3  & 0.34&  1.2\cdot10^6 & 121 & -0.21  & 0.42 & 0.038 & 1.52 & 0.53 & 0.19 & $vf$ \\ 
D6  & 512^3  & 0.34&  1.2\cdot10^6 & 250 & -0.20  & 0.40 & 0.039 & 1.42 & 0.51 & 0.20 & $vf$ \\ 
\hline
D2d & 128^2\times192  & 1.37 & 3.1\cdot10^5 & 23  & -0.27  & 0.54 & 0.029 & 2.44 & 0.89 & 0.08 & $vf$ \\ 
D2w & 256^2\times128  & 1.37 & 3.1\cdot10^5 & 26  & -0.24  & 0.48 & 0.033 & 1.56 & 0.68 & 0.15 & $vf$ \\ 
\hline
E1  & 128^3  & 1.37 & 3.1\cdot10^5 & 42  & -0.03  & 0    & 0.052 & 0.03 & 0.08 & 0.64 & $vf$ \\ 
E2  & 128^3  & 1.37 & 3.1\cdot10^5 & 33  & -0.08  & 0    & 0.041 & 0.83 & 0.48 & 0.19 & $vf$ \\ 
E3  & 128^3  & 1.37 & 3.1\cdot10^5 & 30  & -0.15  & 0    & 0.038 & 1.08 & 0.63 & 0.13 & $vf$ \\ 
E4  & 128^3  & 1.37 & 3.1\cdot10^5 & 29  & -0.22  & 0    & 0.036 & 1.14 & 0.74 & 0.09 & $vf$ \\ 
E5  & 128^3  & 1.37 & 3.1\cdot10^5 & 41  & -0.22  & 0    & 0.051 & 1.01 & 0.55 & 0.14 & $vf$ \\ 
\hline
VSA & 128^3  & 1.37 & 3.1\cdot10^5 & 46  & -0.27  & 0    & 0.058 & 0.27 & 0.07 & 0.90 & $p$  \\ 
VSA'& 128^3  & 1.37 & 3.1\cdot10^5 & 45  & -0.28  & 0    & 0.056 & 0.31 & 0.39 & 0.17 & $nf$ \\ 
           \hline
         \end{array}
     $$ 
\end{table*}

Stress-free boundary conditions are used for the velocity,
\begin{equation}
U_{x,z} = U_{y,z} = U_z = 0,
\end{equation}
and either vertical field or perfect conductor conditions for the
magnetic field,
i.e.\ 
\begin{eqnarray}
B_x = B_y &=& 0, \;\; {\rm (Vertical\;Field)} \\
B_{x,z} = B_{y,z} = B_z &=& 0, \;\; {\rm (Perfect\;Conductor)}
\end{eqnarray}
respectively.
We may think of them as open and closed boundaries, respectively,
because they either do or they do not permit a magnetic helicity flux.
In the $y$ and $x$ directions we use periodic and shearing-periodic
boundary conditions, respectively.
The simulations were made with the {\sc Pencil Code}%
\footnote{\texttt{http://www.nordita.org/software/pencil-code/}},
which uses sixth-order explicit finite differences in space and third
order accurate time stepping method.
Resolutions of up to $512^3$ mesh points were used.

\begin{figure*}[t]
\centering
\includegraphics[width=.925\textwidth]{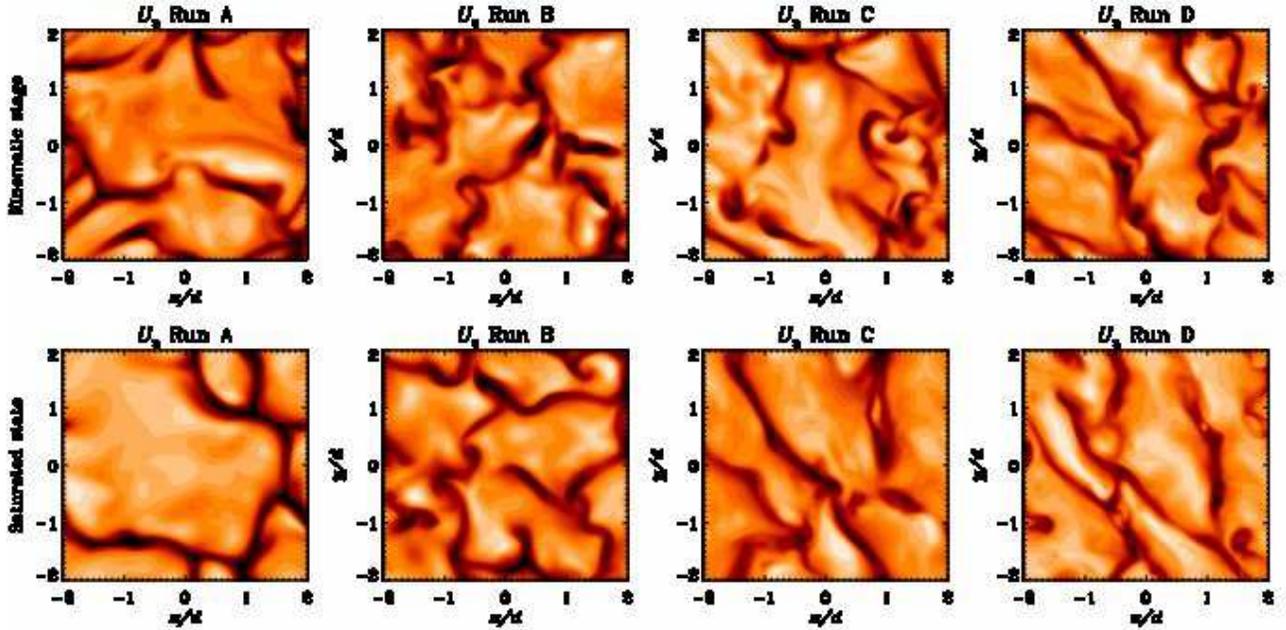}
\caption{
  Vertical velocity in the middle of the convectively unstable layer
  of Runs~A--D (from left to right) during the kinematic
  stage of the simulations [upper row of panels, $t=200\,(d/g)^{1/2}$
  corresponding to $t \urms \kef \approx 50$] and the saturated state
  [lower row, $t=2000\,(d/g)^{1/2}$ or $t \urms \kef \approx 500$].}
\label{fig:pslices}
\end{figure*}

\section{Results}
\label{sec:results}

\subsection{Description of the hydrodynamic state} 

The hydrodynamics of local rotating convection simulations similar to ours 
has been discussed previously in the literature (e.g.\ Brandenburg et al.\ 
\cite{Brandea1996}; K\"apyl\"a et al.\ \cite{Kaepylaeea2004}). 
However, the imposed shear flow used in the present 
runs is a new element. Thus it is interesting to consider the
differences to the cases where it is absent. We consider
four representative runs where physical ingredients are added
individually in order to see their respective effects. In Run~A
neither rotation nor shear are present, whereas in Runs~B and C only 
rotation and only shear are applied, respectively. In Run~D both
shear and rotation are used (see Table \ref{tab:onlyshear} for details).

All our runs start with a weak magnetic field ($10^{-5}$ times the
equipartition value).
For Runs~A to D it takes about 200 turnover times for this field to
grow to values such that it affects the flow,
so we consider this as the kinematic stage of the simulation.
We use this time span to characterize the basic hydrodynamic state.

The upper row of panels in Fig.~\ref{fig:pslices} shows the 
vertical velocity in the middle of the convectively unstable layer
for Runs~A to D.
In the cases without shear (Runs~A and B), the convective pattern is 
dominated by large cells which are isotropic in the horizontal 
plane. In the case with rotation (Run~B), the downflows at the vertices
of convective cells tend to exhibit vortical structures, contributing to
net helicity. When shear is added
the convective cells are slightly elongated along the $y$ direction,
which is the direction of the imposed large-scale flow.

\begin{figure}[t]
\centering
\includegraphics[width=0.95\columnwidth]{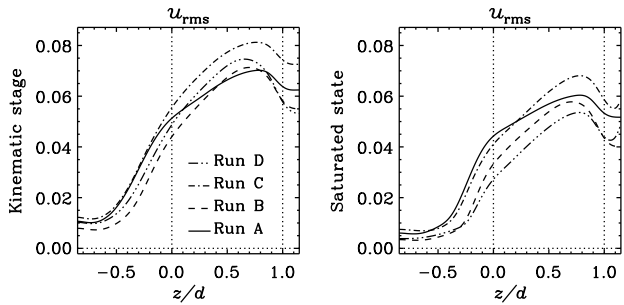}
\caption{Horizontal averages of the rms velocity in the kinematic
  (left panel) and saturated states (right panel) for Runs~A to
  D. Linestyles are indicated in the legend in the left panel. 
  The vertical lines at $z=(0,1)d$ denote the base and top of the 
  convectively unstable layer, respectively.}
\label{fig:purmsz}
\end{figure}

The horizontally averaged root mean square velocity shows
a very similar profile in all models, i.e.\ peaking near the top
of the convectively unstable layer and decreasing towards the base
of the layer, see Fig.~\ref{fig:purmsz}. The runs without shear
are close to each other, with somewhat smaller velocities in 
the case with rotation (Run~B). When a horizontal shear flow is added,
the rms velocity is somewhat larger, but in Run~D $\urms$ decreases again. The shear flow
itself does not contribute to the definition of the rms velocity.

\begin{figure}[t]
\centering
\includegraphics[width=0.95\columnwidth]{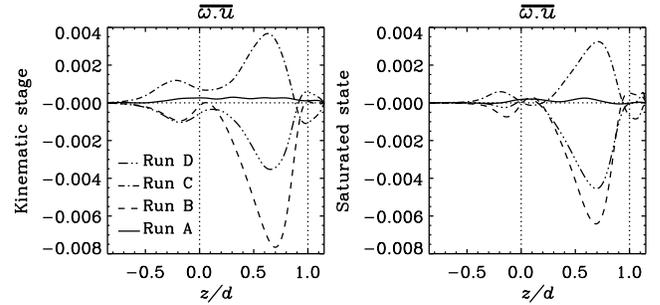}
\caption{Horizontal averages of kinetic helicity in the kinematic
  (left panel) and saturated states (right panel) for Runs~A to
  D. Linestyles are indicated in the legend in the left panel.}
\label{fig:pouz}
\end{figure}

The left panel of Fig.~\ref{fig:pouz} shows the horizontal 
averages of kinetic helicity density for Runs~A to D averaged over the
initial kinematic stage of the simulations. In the case without
rotation or shear (Run~A) very little net helicity is produced. Adding 
rotation corresponding to the north pole ($\vec{g}\cdot\vec{\Omega}<0$; Run~B)
produces significant negative helicity within the convection zone, 
suggestive of a positive $\alpha$-effect. In the case with only 
shear ($S<0$, so  that $\vec{g}\cdot\vec{\nabla}\times\overline{\vec{U}}>0$; Run~C)
the helicity profile is similar to Run~B,
but of opposite sign and roughly half
the magnitude. If both rotation and shear are present, as in 
Run~D, the sign and profile are very similar to Run~B, but the 
magnitude is smaller by approximately a factor of two.
This suggests that there is partial cancellation of the two effects.

\begin{figure}[t]
\centering
\includegraphics[width=0.95\columnwidth]{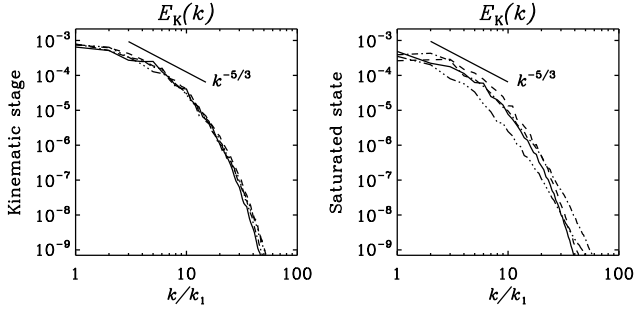}
\caption{Power spectra of velocity from the middle of the
  convectively unstable layer from the kinematic (left panel)
  and saturated states (right panel).
  Linestyles as in Fig.~\ref{fig:purmsz}.
  The straight lines with a slope of $-5/3$ are given for comparison.}
\label{fig:pspecu}
\end{figure}

In the late stages of the simulations when the magnetic field
has saturated the convective pattern does not change significantly
in the cases without shear (see the lower row of 
Fig.~\ref{fig:pslices}). In Runs~C and D, there seems to be 
a larger change,
i.e.\ clearer elongation of cells along the $y$ direction,
possibly due to the strong large-scale magnetic
field that develops in those cases. 
Figures~\ref{fig:purmsz} and \ref{fig:pouz} show that velocity and helicity decrease somewhat
in comparison to the hydrodynamic state in all cases except Run~D where the helicity seems to increase.
Part of this change can be attributed to the fact that some of the runs
were not yet fully relaxed during the kinematic phase and that they are
still undergoing a slow thermal adjustment,
especially in the cases without shear. However, in Runs~C and D 
where the dynamo grows to super-equipartition strengths, the magnetic 
field is likely to be responsible for much of the change.

In the absence of shear the spectra are similar to those published
previously (e.g.\ Brandenburg et al.\ \cite{Brandea1996}), but with 
shear we should really only look at one-dimensional spectra in the
streamwise direction, because this is the only periodic direction.
In the kinematic stage little difference is seen between the four 
runs (see the left panel of Fig.~\ref{fig:pspecu}). However, in the 
saturated state the spectra for the shearing runs are steeper than for
the non-shearing runs.
This may indicate the development of large-scale ordered velocity structures.

\subsection{Excitation of dynamo action}

\begin{figure}[t]
\centering
\includegraphics[width=.9\columnwidth]{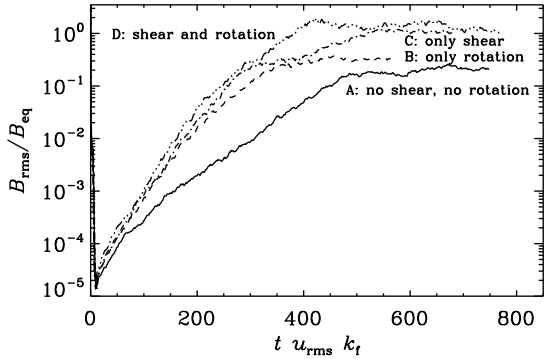}
\caption{Growth of the total rms magnetic field from four runs, which
  were performed without shear or rotation (solid line, Run~A),
  with only rotation (dashed, Run~B), only shear (dash-dotted, Run~C), and
  with shear and rotation (triple-dot-dashed, Run~D), respectively. $\Pm =
  5$, $\rm Rm \approx 60\ldots90$, and grid resolution $256^3$ in all runs,
  see also Table \ref{tab:onlyshear}.}
\label{fig:brms}
\end{figure}

We consider first the effects of shear and rotation on the dynamo.
In Fig.~\ref{fig:brms} we show the resulting growth of the
rms magnetic field for $\Rm=56\ldots74$ and $\Pm=5$
for Runs~A to D introduced in the previous section.
The slowest growth occurs in Run~A with no rotation and no shear
($\lambda=0.016\,\urms\kef$).
Adding rotation (Run~B; $\Co=0.42$) almost doubles the growth rate
($\lambda=0.03\,\urms\kef$), and the saturation field strength
of the horizontally averaged field
is somewhat higher: $\overline{B}\approx0.4\,B_{\rm eq}$.
Next, adding shear (Run~C) raises the growth rate further and,
more importantly, it raises the saturation field strength by a factor
of about 3 ($\overline{B}\approx1.1\,B_{\rm eq}$).
Again, adding rotation (Run~D) increases the growth rate further
($\lambda=0.04\,\urms\kef$), 
and increases the saturation level by almost fifty per cent.
In the absence of shear and rotation the critical Reynolds number is 
$\approx30$ for open (vertical field) boundary conditions 
(see Sect.~{\ref{sec:bcs}} for more details on the effects of
boundary conditions).
If shear is included, the dynamo is excited already at 
$\Rm\approx5$.

We note that in the kinematic phase the mean and total fields have the
same growth rate.
The two begin to depart from each other only in the saturation phase
which is achieved more quickly for the small-scale field.
This behaviour is reminiscent of dynamos where the saturation of the
large-scale field is controlled by magnetic helicity evolution
(Blackman \& Brandenburg \cite{BB02}).

\begin{figure}[t]
\centering
\includegraphics[width=0.925\columnwidth]{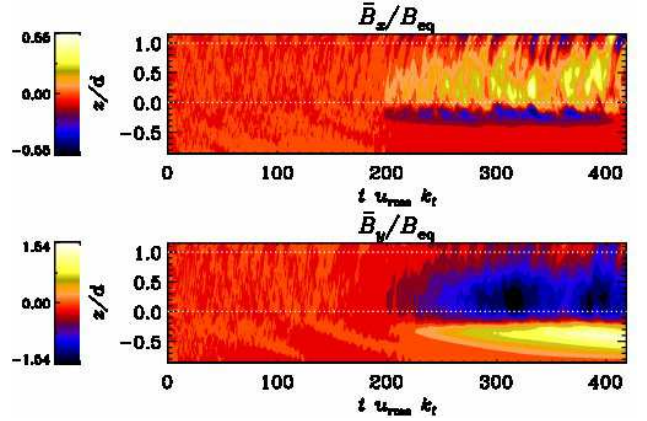}
\caption{Horizontally averaged magnetic fields $\mean{B}_x$ (upper
  panel) and $\mean{B}_y$ (lower panel),
  normalized by the volume average of $\Beq$,
  as functions of time and $z$
  for Run~D5 with ${\rm Rm} = 121$, $\Sh=-0.21$, and $\Co=0.42$.
  The dotted white lines show top ($z=d$) and bottom
  ($z=0$) of the convection zone.}
\label{fig:but256a}
\end{figure}

\begin{figure}[t]
\centering
\includegraphics[width=0.925\columnwidth]{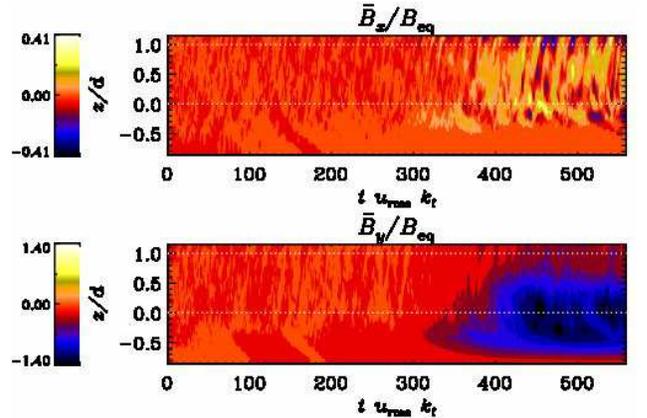}
\caption{Same as Fig.~\ref{fig:but256a}, but for Run~C with $\Rm=74$,
  $\Sh=-0.17$, and $\Co=0$.}
\label{fig:but_norot}
\end{figure}

\begin{figure}[t]
\centering
\includegraphics[width=0.925\columnwidth]{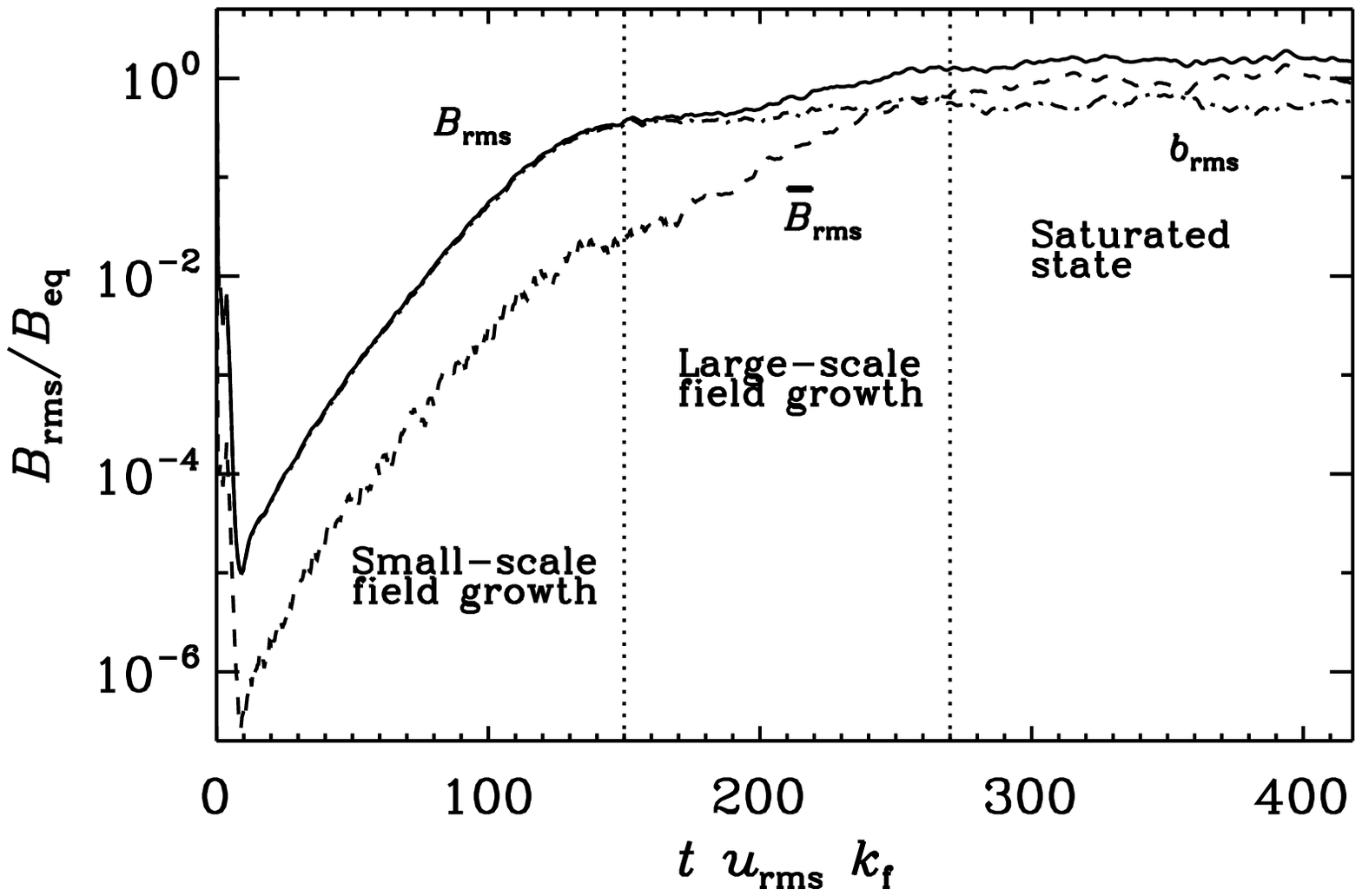}
\caption{Rms-values of the total magnetic field (solid line), the
mean field $\mean{B}_{\rm rms}=\brac{\mean{B}_x^2+\mean{B}_y^2}^{1/2}$
(dashed line) and fluctuating field with
$b_{\rm rms}^2 = \brms^2 - \mean{B}^2_{\rm rms}$
(dot-dashed line) for Run~D5.}
\label{fig:totalmean}
\end{figure}

\begin{figure*}[t]
\centering
\includegraphics[width=.875\textwidth]{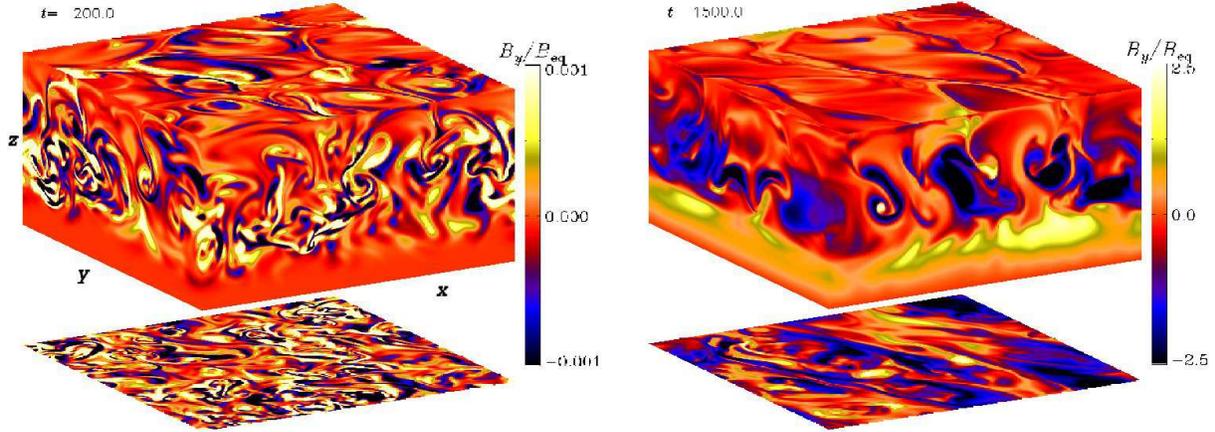}
\caption{Snapshots of $B_y$ in the early phase ({\it left}:
  $t=200\,(d/g)^{1/2}$, corresponding to $t \urms \kef \approx 50$) and
  saturated phase ({\it right}: $t=1500\,(d/g)^{1/2}$, or $t \urms \kef \approx
  360$) for Run~D5. The sides of the box show the periphery of the
  domain whereas the top and bottom slices show $B_y$ at vertical
  heights $z=d$ and $z=0$, respectively. See also
  \texttt{http://www.helsinki.fi/\ensuremath{\sim}kapyla/movies.html}}
\label{fig:By256b}
\end{figure*}

\subsection{Large-scale magnetic fields}
\label{LSfields}

Adding a large-scale shear flow not only makes the dynamo easier to
excite, it also helps the generation of large-scale magnetic
fields and, most importantly, it is expected to drive magnetic helicity fluxes
along lines of constant shear
(Vishniac \& Cho \cite{VishCho2001}; Subramanian \& Brandenburg \cite{SB06}).
A representative solution from Run~D5 with ${\rm Rm}=121$,
$\Sh = -0.21$, and $\Co = 0.42$ is shown in Fig.~\ref{fig:but256a}.
Note that in runs with rotation the large-scale field has opposite
sign in the overshoot layer and the convection zone, whereas in models
with only shear the field has the same sign everywhere;
see Fig.~\ref{fig:but_norot}.

\begin{figure}
\centering
\includegraphics[width=0.925\columnwidth]{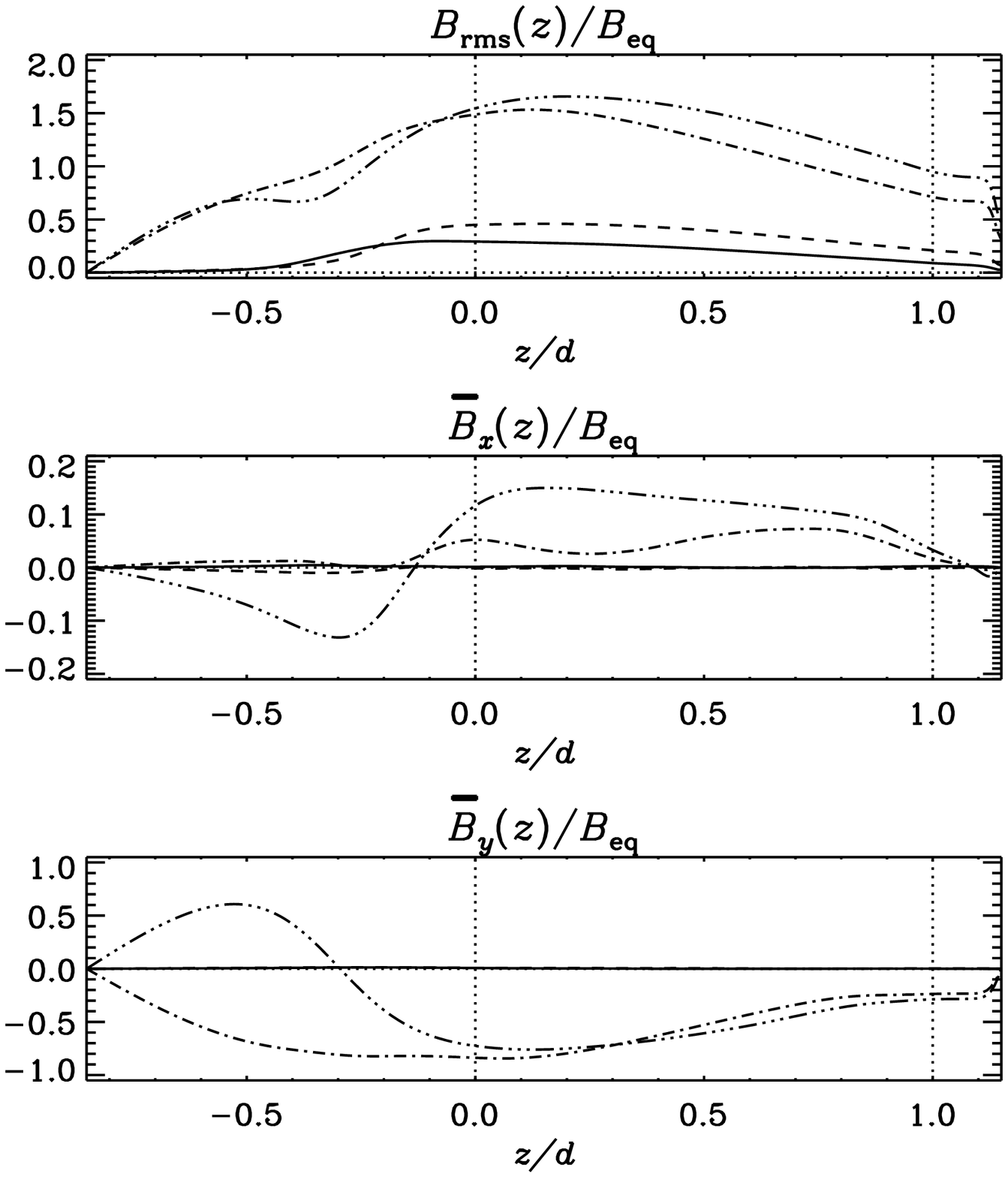}
\caption{Horizontal averages of the total rms magnetic field
  (top panel), $\mean{B}_x(z)$ (middle), and $\mean{B}_y(z)$ (bottom)
  for Runs~A to D. The data are averaged in time over the
  saturated state of the dynamo. Linestyles as indicated in
  Fig.~\ref{fig:purmsz}. The dotted vertical lines denote the
  convectively unstable layer between $0 < z < d$.}
\label{fig:pmfield}
\end{figure}

In all cases the magnetic energy grows exponentially from a weak
small-scale seed magnetic field.
Generally the small-scale dynamo saturates first and
a large-scale field develops later; see Fig.~\ref{fig:totalmean}
for representative results from Run~D5.
Figure~\ref{fig:By256b} shows that during this process the magnetic
field changes from a filamentary field to a more diffuse one.
This behaviour has also been seen in simulations of forced turbulence
with shear and open boundaries (Brandenburg 2005).
The signs of $\mean{B}_x$ and 
$\mean{B}_y$ are always opposite to each other; see Fig.~\ref{fig:pmfield}.
This is simply because of negative shear,
$S<0$, so, for example, a positive $\mean{B}_x$
results in a negative $\mean{B}_y$.
In some simulations we 
have observed sign changes of the large-scale field in the saturated
stage but these do not seem to be periodic but rather one-off events.

The rms value of the mean field contains a major fraction of the total field, i.e.\
$\brac{\mean{B}_y^2}^{1/2}/\brms \approx 0.5 \ldots 0.8$.
Without shear the large-scale field is weaker
by an order of magnitude (Table \ref{tab:onlyshear}).
Note that at comparable Reynolds numbers
the small-scale dynamo is excited in the absence of rotation
and shear (Run~A). A very similar
large-scale
field pattern is also obtained for ${\rm Rm} \approx 10$ and
$\approx30$ where the small-scale dynamo is absent or marginally excited.
Further below we will show that the final saturation levels are only
weakly dependent on $\Rm$.

\begin{figure}[t]
\centering
\includegraphics[width=0.95\columnwidth]{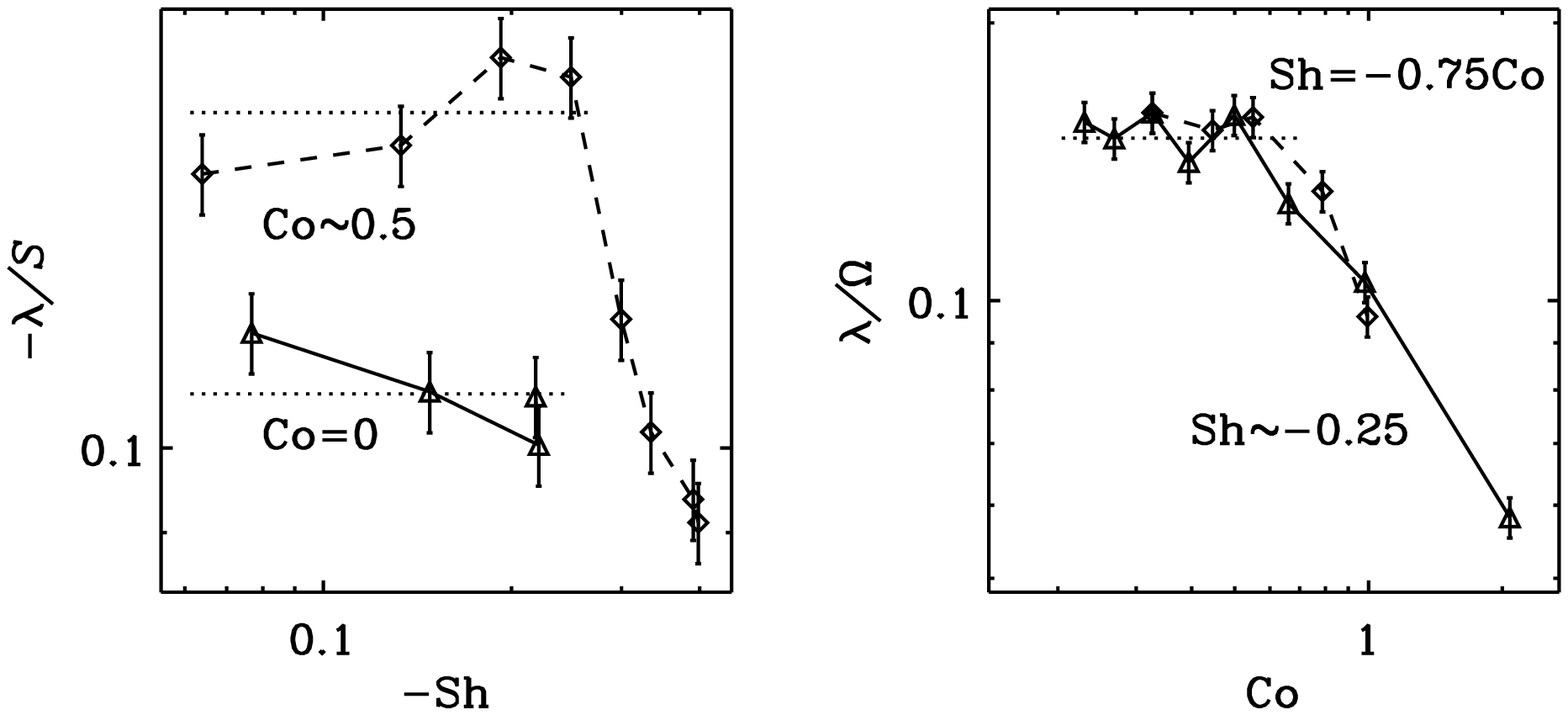}
\caption{
Left panel: growth rate of $\brac{\mean{B}_y^2}^{1/2}$ divided by the
shear rate $S$ as a function of $\Sh$ for two set of runs with $\Co=0$ 
(solid line) and $\Co\approx0.5$ (dashed line). Right panel: growth 
rate of $\brac{\mean{B}_y^2}^{1/2}$ normalized by the 
rotation rate $\Omega$ for two set of runs with $\Sh\approx-0.25$
(solid line) and $\Sh=-\threefourths \Co$ (dashed line). 
${\rm Rm} \approx 30$ in all runs. The horizontal dotted lines show 
curves proportional to the shear rate $S$ (left panel) and rotation rate $\Omega$ (right panel) for reference.}
\label{fig:grRm30}
\end{figure}

It turns out that in the range $0.05<|\mbox{Sh}|<0.25$
the growth rate is proportional to the shear rate,
so the ratio $\lambda/|S|\approx0.1$ is approximately independent of
the value of $|S|$; see Fig.~\ref{fig:grRm30}, where
$\lambda$ is the growth rate of $\brac{\mean{B}_y^2}^{1/2}$.
The same scaling was found in the simulations of
non-helical turbulence with shear (Yousef et al.\ \cite{Yousefea2008a}).
It becomes increasingly difficult to perform runs with $-\Sh > 0.2$ 
without rotation due to the simultaneous generation of large-scale 
vorticity. The theory for such a `vorticity dynamo' has been developed
by Elperin et al.\ (\cite{Elperinea2003}) and has also been found 
numerically from forced turbulence simulations by 
Yousef et al.\ (\cite{Yousefea2008a,Yousefea2008b})
and Brandenburg et al.\ (\cite{Brandea2008}).
The vorticity generation is an interesting subject in itself and has been
studied in a separate paper (K\"apyl\"a et al.\ \cite{KMB09}). To avoid
complications due to the vorticity dynamo we add rotation into the 
system which stabilizes the shear terms in the Navier-Stokes 
equations.

When rotation is added, the growth rate divided by the shear rate
seem to stay approximately constant for $\Sh>-0.25$. When shear is
increased further, $-\lambda/S$ decreases rapidly, see the left panel
of Fig.~\ref{fig:grRm30}. On the other hand, if the shear is kept 
constant the quantity $\lambda/\Omega$ stays constant up to 
$\Co\approx0.5$ after which it decreases rapidly, see the right panel
of Fig.~\ref{fig:grRm30}. The last case considered is to fix
the ratio $q\equiv-S/\Omega=1.5$. In that case the
behaviour of $\lambda/\Omega$ is almost identical to the case where
$\Sh$ was kept fixed. A very similar trend was found by Yousef et 
al.\ (\cite{Yousefea2008b}).

\begin{figure}[t]
\centering
\includegraphics[width=0.95\columnwidth]{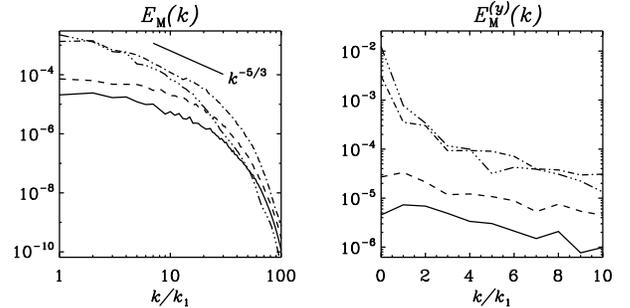}
\caption{One-dimensional power spectra of the magnetic field
  $E_{\rm M}(k)$ (left panel) and the $y$-component of the magnetic
  field $E^{(y)}_{\rm M}(k)$ (right panel) in the saturated state of the dynamo. Linestyles as indicated in
  Fig.~\ref{fig:purmsz}.
  The straight lines with a slope of $-5/3$ are given for comparison.}
\label{fig:pspecb}
\end{figure}

The left hand panel of Fig.~\ref{fig:pspecb} shows one dimensional power
spectra of the magnetic field in the saturated state for Runs~A 
to D. The runs with shear exhibit a steeper spectrum in comparison to
the cases without shear.
However, it is
clear, that the spectrum of the streamwise component of the magnetic field
is sharply peaked at $k_y=0$ and drops almost by an order of magnitude at the
next wavenumber 1 (in units of $k_1$), because of the large scale field.
This is the reason why we have also plotted spectra in linear scale
in the right hand panel so
as to accommodate the $k_y=0$ wavenumber.

\subsection{Dependence on box size}
\label{BoxSize}
\begin{figure}
\centering
\includegraphics[width=0.925\columnwidth]{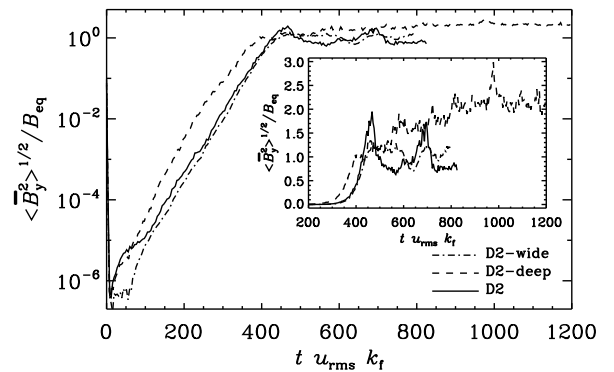}
\caption{The rms value of $\mean{B}_y$ for the runs
  indicated in the legend. The inset shows the same in linear scale.}
\label{fig:p128b}
\end{figure}

Convective overshooting in the present simulations is comparable
to that presented in K\"apyl\"a et al.\ (\cite{Kaepylaeea2004}) and
certainly does not reach all the way down to the lower boundary.
However, in order to determine whether the lower boundary
still affects the results appreciably, we modify
Run~D2 such that the lower stable 
layer reaches down to $z=-1.85d$ in 
comparison to $z=-0.85d$ in the other runs.
We denote this run as D2--deep.
We find that the 
large-scale magnetic field develops much the same as in the 
standard case, with the exception that the saturation of the field in the overshoot
layer is slower, see Fig.~\ref{fig:p128b}. In the saturated state the magnetic 
field occupies the whole overshoot layer.
This is probably due to the large diffusion and the shear flow
in the overshoot layer which are absent in the Sun.

Figure \ref{fig:pslices} shows that relatively few convective 
cells are present in the computational domain. In order to test
the effects of increasing the horizontal size of the domain we
again take Run~D2 as the basis and double the horizontal extent.
This run is referred to as D2--wide.
We find that the average magnetic field shows less fluctuations.
This is due to the averaging over a larger number of convective cells. 
Growth rate and saturation level of the magnetic
field are slightly larger than those of Run~D2, see Fig.~\ref{fig:p128b}.
This is because a bigger box results in more scale separation,
which in turn increases the mean-field dynamo effect;
see Eq.~(80) of Brandenburg et al.\ (2002).

\begin{figure}[t]
\centering
\includegraphics[width=0.925\columnwidth]{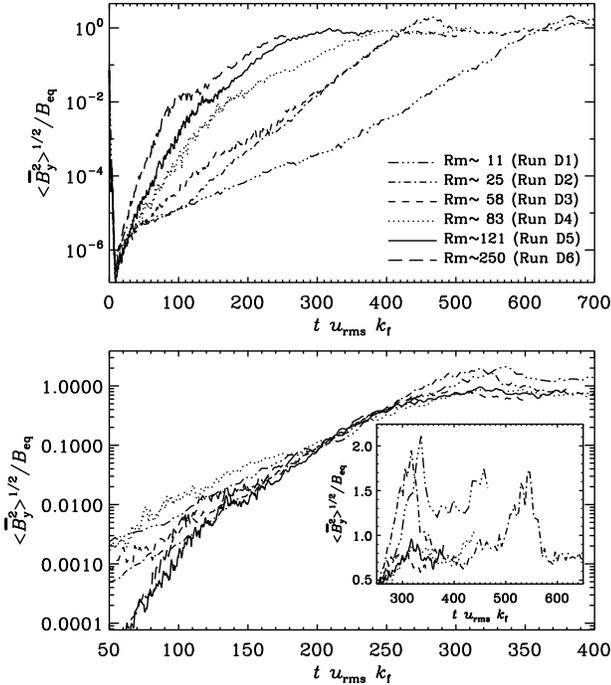}
\caption{Upper panel: root mean square values of $\mean{B}_y$ for the
  Runs~D1 to D6. Linestyles as indicated in the legend. Lower panel:
  close-up of the interval where the dynamo saturates for the five
  largest Reynolds numbers. Data for the lower Reynolds numbers are
  shifted so that the saturation occurs approximately at the same
  point in the figure. The inset shows the saturated state in linear
  scale.}
\label{fig:pcomp_gr}
\end{figure}

\subsection{Dependence on $\Rm$}
Originally catastrophic quenching of the mean-field dynamo 
effect was conjectured to derive from the action of the 
small-scale magnetic field whose magnitude was expected to increase
as the magnetic Reynolds number is increased (Vainshtein 
\& Cattaneo \cite{VainCatt1992}). This 
should lead to $\Rm$-dependent saturation levels of the magnetic field.
In our Runs~D1 to D6 listed in Table \ref{tab:onlyshear}, $\Rm$ is 
varied from 11 (roughly a third of the critical value for exciting the 
small-scale dynamo) to 250 (roughly eight times supercritical).
In all of these runs the saturation level of the magnetic field is
essentially the same, bearing in mind the large fluctuations
in the relatively short time series of the higher resolution runs;
see Fig.~\ref{fig:pcomp_gr}.
Below, we demonstrate that the boundary conditions play a much
more crucial role in allowing large-scale dynamos to operate
efficiently.

\begin{figure}[t]
\centering
\includegraphics[width=0.925\columnwidth]{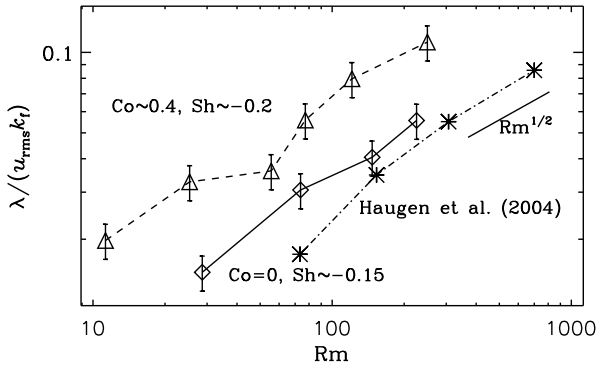}
\caption{Growth rate of the total magnetic field for runs with
only shear (solid line) and Runs~D1 to D6 with both shear and rotation 
(dashed line).}
\label{fig:pgrrm}
\end{figure}

We recall that our definition of $\kef$ in terms of the depth of the
convectively unstable layer ($2\pi/d$) is somewhat arbitrary.
However, it turns out that our values of $\lambda/(\urms\kef)$ are in
agreement with those of forced turbulence (Haugen et al.\ \cite{Haugenea2004}),
where $\kef$ is well defined.
Over the range $35<\Rm<250$, our values of $\lambda$ scale still almost linearly
with $\Rm$ (see Fig.~\ref{fig:pgrrm}); it is known from forced
turbulence simulations that the $\Rm^{1/2}$ scaling occurs only for $\Rm>200$.

The growth rate of the mean field in the late stages of growth 
does not seem to be strongly dependent on $\Rm$, see the lower 
panel of Fig.~\ref{fig:pcomp_gr}.
Note also that the saturation field strengths are similar for different
values of $\Rm$.

\subsection{Effects of boundary conditions}
\label{sec:bcs}
\begin{figure}[t]
\centering
\includegraphics[width=0.925\columnwidth]{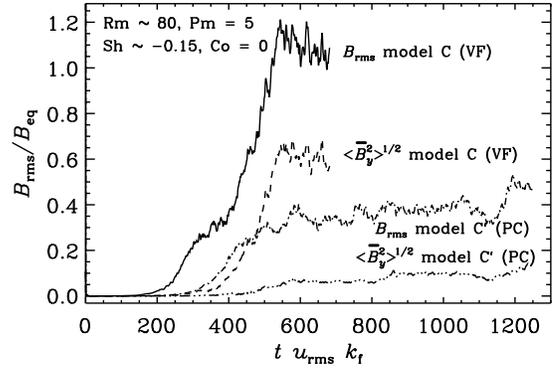}
\caption{Root mean square values of the total magnetic field and the
  horizontally averaged component $\mean{B}_y$ for Runs~C and
  C'. VF denotes vertical field and PC perfect conductor boundary
  conditions.}
\label{fig:en256a}
\end{figure}

Here we compare Runs~C and C' made with vertical field and perfect
conductor boundary conditions, respectively. The former allows flux of magnetic
helicity across the boundary whereas the latter does not.
Figure~\ref{fig:en256a} shows the time evolution of $\brms$ and
$\brac{\mean{B}_y^2}^{1/2}$ for these runs.
The run with vertical field boundary conditions 
clearly exhibits dynamically important large-scale magnetic fields. The
rms value of the \emph{mean} magnetic field is approximately 60 per cent of the
equipartition value of the turbulent velocity field (see also Table
\ref{tab:onlyshear}). Contrasting this with the same run performed
with perfect conductor boundaries, the difference is striking:
although the total magnetic field is roughly 40 per cent of the equipartition value by
virtue of small-scale dynamo action, the mean field is weak,
contributing now only roughly 20 per cent of the total.

The difference between the runs can be understood in terms of magnetic
helicity evolution: allowing a non-zero vertical field at the
boundary enables small-scale magnetic helicity to escape from the
computational domain.
With perfect conductor boundary conditions the magnetic helicity flux
vanishes and thus the total magnetic helicity can
change only resistively, which leads to slow saturation of the large-scale field.
The fact that large-scale magnetic fields have not yet been seen
in convection simulations
with perfectly conducting boundaries does therefore not mean that this is 
not possible, but rather that it would presumably take a very long time
(see the slow but persistent rise for model~C' in Fig.~\ref{fig:en256a}).

\subsection{Horizontal versus vertical shear}

In an earlier study, Tobias et al.\ (\cite{Tobiasea2008}) used a vertical 
shear
profile, i.e.\ $\mean{U}_y^{(0)}(z)$,
and found that no appreciable large-scale magnetic fields
were generated
even though open boundary conditions were used in the $z$-direction.
In order to compare with their results we have made 
runs with a shear profile
\begin{equation}
\meanv{U}^{(0)} = \onehalf U_0 \bigg[1 + \tanh\Big(\frac{z-z_2}{d_1}\Big)\bigg] \hat{\bm{e}}_y \;,
\end{equation}
where $U_0=-0.1 (gd)^{1/2}$ and $d_1=0.4d$. This flow is imposed through 
an additional relaxation term in the Navier--Stokes equation of the form
\begin{equation}
\pd_t \bm{U} = \ldots -\tau^{-1} \left(\bm{U}-\meanv{U}^{(0)}\right),
\end{equation}
where $\tau=0.5\,(d/g)^{1/2}$.
In these runs with vertical shear we put $S=0$.

\begin{figure}
\centering
\includegraphics[width=0.925\columnwidth]{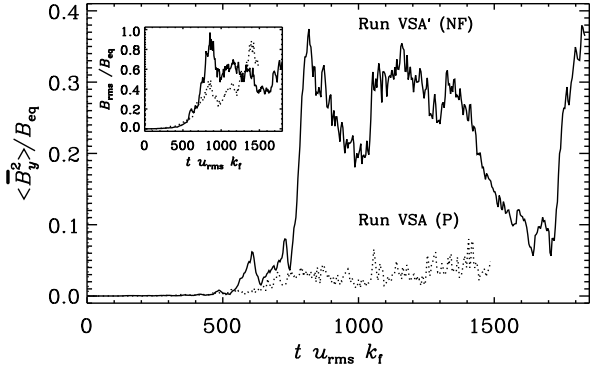}
\caption{
  Root mean square values of the horizontally averaged component
  $\mean{B}_y$ for Runs~VSA' (solid line) and VSA (dotted line). The
  inset shows the rms values of the total magnetic field from the same
  runs.
  NF denotes normal field condition and P periodic boundary condition in
  the $x$-direction.}
\label{fig:pvs}
\end{figure}

\begin{figure}
\centering
\includegraphics[width=0.925\columnwidth]{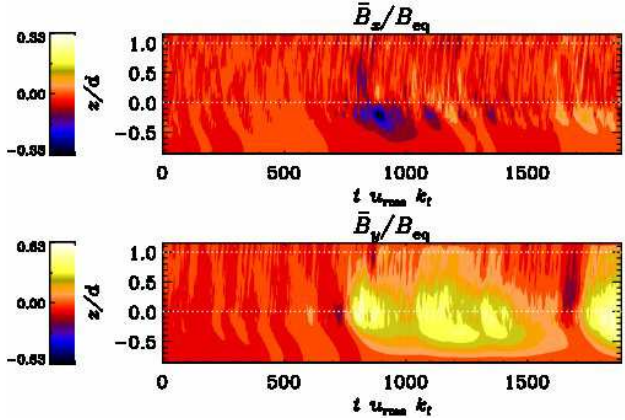}
\caption{Horizontally averaged magnetic fields $\mean{B}_x$ (upper
  panel) and $\mean{B}_y$ (lower panel),
  normalized by the volume average of $\Beq$,
  as functions of time and $z$ for Run~VSA'.}
\label{fig:st_VSap}
\end{figure}

We confirm the results of Tobias et al.\ (\cite{Tobiasea2008}) in the 
case where the horizontal boundaries are periodic and find that the 
rms value of the large-scale (horizontally averaged) field 
amounts to only a few per cent of the total field. We believe that
this is due to the inefficiency of the magnetic helicity flux through
the vertical boundaries. In this case the isocontours of shear are 
horizontal and thus the shear-mediated flux, which we conjecture to 
be efficient, cannot leave the system. To test this hypothesis, we have made
additional runs where the $x$-direction is no longer periodic but instead a
stress-free boundary condition for velocity and a normal-field
condition for the magnetic field according to
\begin{eqnarray}
U_x = U_{y,x} = U_{z,x} = B_y = B_z = 0
\end{eqnarray}
is used.
In this case the shear-mediated magnetic helicity flux can escape through the
$x$-boundaries
and significant (up to 50 per cent of the total)
large-scale magnetic fields are indeed produced, see
Fig.~\ref{fig:pvs}. This result lends credence to the hypothesis 
that the shear-induced flux is an important ingredient in 
allowing large-scale dynamos to saturate
to full strength on a dynamical timescale.

In order to illuminate the circumstances around the time when the mean
toroidal field becomes suddenly quite weak ($t\urms\kef\approx1600$),
we show in Fig.~\ref{fig:st_VSap} the mean fields in the cross-stream
and streamwise directions.
Note that there is no field reversal associated with the sudden drop of
toroidal magnetic field, as one might have expected from simulations of
dynamo action that is driven purely by magnetic buoyancy, as in the work
of Cline et al.\ (\cite{Clineea2003}).

\subsection{Surface appearance of large scale field}

\begin{figure}
\centering
\includegraphics[width=0.925\columnwidth]{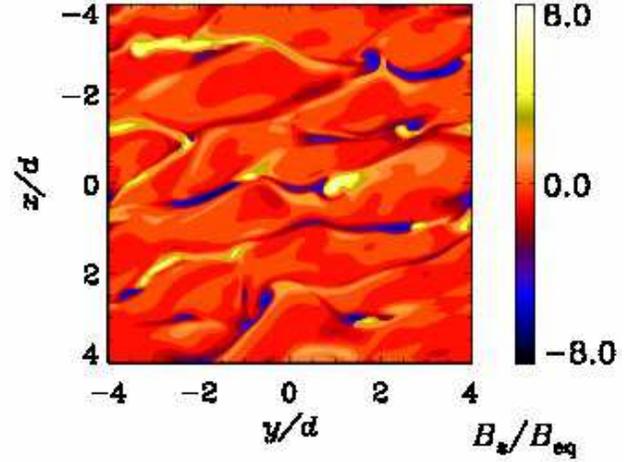}
\caption{Vertical magnetic field component at the top of the convectively unstable
layer at $z=z_3$ for Run~D2--wide toward the end of the run.
The plot is oriented such that the mean shear flow points to the
left in the lower part of the plot ($x/d=4$) and to the right in the
upper part of the plot ($x/d=-4$). See also
\texttt{http://www.helsinki.fi/\ensuremath{\sim}kapyla/movies.html}}
\label{fig:Bzslices}
\end{figure}

In order to make contact with observations, one must eventually compute
the field as it would be observable at the surface of the domain.
This process can be rather complicated and would involve radiation
transfer of polarized light.
Here we consider instead just the vertical components of the magnetic
field at the top of the convectively unstable layer at $z=z_3$;
see Fig.~\ref{fig:Bzslices}.

Two things are immediately evident: firstly, the magnetic structures
tend to be inclined by up to 30 degrees relative to the $y$-direction
and, secondly, there are
bipolar regions with a systematic magnetic field orientation such
that a negative polarity follows a positive one, which is in agreement
with an overall positive mean toroidal field.
Comparing with a similar plot of Brandenburg (\cite{Brandenburg2005}),
the bipolar regions are here much more compact.
Also, the inclination is here such that it would correspond to the southern
hemisphere, because $\mbox{d}\overline{U}_y/\mbox{d}x$ is here negative.


\section{Conclusions}
\label{sec:conclusions}

The present simulations provide a clear demonstration
that convection can produce large-scale magnetic fields of equipartition
strengths in an open Cartesian domain with shear crossing the surface.
Our results also demonstrate that this is not
possible within a comparable time span using
a closed domain with perfectly conducting boundaries.
The sensitivity on boundary conditions has also been noted in earlier
work with forced turbulence and shear (Brandenburg \cite{Brandenburg2005}).
This can be interpreted as being due to magnetic helicity fluxes
allowing the system to dispose of excess small-scale magnetic helicity
along contours of constant shear
(Vishniac \& Cho \cite{VishCho2001}; Brandenburg \& Subramanian \cite{BS05b};
Subramanian \& Brandenburg \cite{SB06}).
In our case this is the vertical direction, but in
Tobias et al.\ (\cite{Tobiasea2008}), it is the horizontal direction.
Thus, in our case shear-driven magnetic helicity flux can escape the domain whereas in theirs
it cannot due to the periodicity in the horizontal directions. This could
explain why no large-scale dynamo develops in their case.
Their model is otherwise similar to ours (same values of $\Co$ and $\Pm$,
whilst $\Rm$ is about two times smaller, and $\Sh$ is twice as large.)
Control simulations show that replacing their periodic boundary
condition in the $x$ direction by a normal-field boundary condition
allows strong large-scale fields also in their setup (Fig.~\ref{fig:pvs}).

The precise nature of the dynamo found here cannot be pinned
down rigorously unless one is able to identify unambiguously the mechanisms
that are actively at work.
In addition to the $\alpha$-effect, there are other possible
alternatives, in particular the incoherent alpha-shear dynamo
(Vishniac \& Brandenburg \cite{VishBran1997}; Proctor \cite{Proctor2007}),
which could produce a large-scale magnetic field even without
stratification and hence no regular $\alpha$-effect.
This interpretation has been favoured in some recent work where the
shear--current effect was found not to be excited
(Brandenburg et al.\ \cite{Brandea2008}).
However, this matter is still under debate and there are arguments in favour
of a shear--current effect (e.g.\ Kleeorin \& Rogachevskii \cite{KleRoga2008}).

In order to disentangle the relative roles of different mechanisms in
the present work it will be important to establish a closer connection with
mean-field theory by determining the components of the $\alpha$ and turbulent
magnetic diffusivity tensors for different boundary conditions and as
a function of $\Rm$.
Only when the results are shown to be converged with $\Rm$ may we expect
them to be relevant for the Sun and perhaps other astrophysical bodies.

We emphasize that large-scale dynamo action is normally only seen
in the nonlinear stage, whilst the linear stage is dominated by
small-scale magnetic fields.
It is therefore important to consider mean-field transport coefficients
that are affected by the magnetic field.
It is in principle even possible that the relevant dynamo mechanism
is an intrinsically nonlinear one, as it is in the case of accretion
disc turbulence where the turbulence is the result of the dynamo itself
(Brandenburg et al.\ \cite{Brandea1995}).
The present results do not give any indications in this direction,
because there seems to be a continuous transition from low-$\Rm$ 
to higher-$\Rm$ large scale dynamo action.
For example at $\Rm=11$, small-scale dynamo action in the usual sense
is not expected, and nevertheless, the final saturation strength of the
large-scale field is similar to the case of higher-$\Rm$ dynamos.

Another important aspect to consider in future work is the degree
of scale separation.
As the Reynolds number increases, the turbulence becomes more vigorous
and the convection plumes will no longer penetrate the entire
depth of the unstable layer.
Therefore, we expect the wavenumber of the energy-carrying eddies to
increase, i.e.\ $\kef>2\pi/d$.
This means that the degree of scale separation, i.e.\ the ratio $\kef/k_1$,
will also increase.
However, as discussed at the end of Sect.~\ref{BoxSize}, more scale
separation is only expected to enhance large-scale dynamo action.

\begin{acknowledgements}
  The computations were performed on the facilities hosted by
  CSC - It Center for Science Ltd. in Espoo, Finland, who are administered
  by the Finnish Ministry of Education. This research has greatly benefitted 
  from the computational resources granted by the CSC to the grand 
  challenge project `Dynamo08'. PJK acknowledges the financial
  support from the Academy of Finland grant No.\ 121431. The authors
  acknowledge the hospitality of Nordita during the program
  `Turbulence and Dynamos' during which this work was initiated.
  This research was supported in part by the National Science
  Foundation under grant PHY05-51164.
  The authors wish to thank Igor Rogachevskii, Gunther R\"udiger and 
  Alexander Schekochihin and an anonymous referee
  for their comments on the manuscript.
\end{acknowledgements}

\end{document}